\documentclass{ecai}
\usepackage{latexsym}

\usepackage{microtype}
\usepackage{graphicx}
\usepackage{subfigure}
\usepackage{booktabs} 
\usepackage{float}
\usepackage{caption}

\usepackage{amsmath}
\usepackage{amssymb}
\usepackage{mathtools}
\usepackage{amsthm}

\usepackage[colorlinks, linkcolor=blue]{hyperref}

\begin{document}

\begin{frontmatter}

\title{Multi-agent Cooperative Games Using Belief Map Assisted Training}

\author[A,†]{\fnms{Chen}~\snm{Luo}}
\author[A,†]{\fnms{Qinwei}~\snm{Huang}}
\author[B]{\fnms{Alex B.}~\snm{Wu}}
\author[C]{\fnms{Simon}~\snm{Khan}}
\author[D]{\fnms{Hai}~\snm{Li}}
\author[A]{\fnms{Qinru}~\snm{Qiu}\thanks{Corresponding Author. Email: qiqiu@syr.edu.}}

\address[A]{Department of Electrical Engineering \& Computer Science, Syracuse University}
\address[B]{Fayetteville-Manlius High School}
\address[C]{Air Force Research Laboratory}
\address[D]{Department of Electrical Engineering \& Computer Science, Duke University}
\address[†]{Equal Contribution}

\begin{abstract}
In a multi-agent system, agents share their local observations to gain global situational awareness for decision making and collaboration using a message passing system. When to send a message, how to encode a message, and how to leverage the received messages directly affect the effectiveness of the collaboration among agents. When training a multi-agent cooperative game using reinforcement learning (RL), the message passing system needs to be optimized together with the agent policies. This consequently increases the model’s complexity and poses significant challenges to the convergence and performance of learning. To address this issue, we propose the Belief-map Assisted Multi-agent System (BAMS), which leverages a neuro-symbolic belief map to enhance training. The belief map decodes the agent's hidden state to provide a symbolic representation of the agent's understanding of the environment and other agents' status.  The simplicity of symbolic representation allows the gathering and comparison of the ground truth information with the belief, which provides an additional channel of feedback for the learning. Compared to the sporadic and delayed feedback coming from the reward in RL, the feedback from the belief map is more consistent and reliable. Agents using BAMS can learn a more effective message passing network to better understand each other, resulting in better performance in the game. We evaluate BAMS's performance in a cooperative predator and prey game with varying levels of map complexity and compare it to previous multi-agent message passing models. The simulation results showed that BAMS reduced training epochs by 66\%, and agents who apply the BAMS model completed the game with 34.62\% fewer steps on average.
\end{abstract}

\end{frontmatter}

\section{Introduction}

A multi-agent cooperative game involves multiple autonomous systems collaborating with each other to achieve a common goal and maximize the overall utility of the system. These games can be used to model various applications, such as rescue missions where multiple robots are deployed to search for missing persons, military operations where multiple UAVs survey a large area, and scientific expeditions where rovers explore unknown terrain together. However, as the number of agents increases, centralized monitoring, controlling, and optimization becomes infeasible due to the exponential growth in complexity \cite{huang2015energy}\cite{lowe2017multi}. It will also increase the vulnerability of the system to single-point failures \cite{lynch2009single}\cite{moradi2016centralized}.  To overcome these issues, distributed control and optimization are introduced, where each agent makes its own decisions based on local information. However, this approach also has limitation, as agents only have partial observations of their immediate surroundings, and may not be able to make globally optimal decisions.

Message exchanges among agents can provide global information and help the agents move out of local optima. However, excessive communication can consume communication energy, bandwidth, and processing power. Sending redundant messages in consecutive cycles, or by agents close to each other, is likely to waste resources. Additionally, frequently communicating every piece of observed information can be wasteful and also undermine the receiver’s decision-making ability. Furthermore, to save communication energy and improve security, the high-dimensional observation should be encoded into a low-dimensional message that can only be decoded by the agents. Therefore, when to communicate, what to communicate and how to encode/decode the message are variables that need to be optimized. Reinforcement learning (RL), such as the actor-critic model, is commonly used to optimize multi-agent games. Manually design message passing system usually does not work well with the RL due to the lack of prior knowledge of the features that are needed by the policy network.  A typical approach \cite{das2019tarmac}\cite{singh2018learning} is to train the message passing network together with the policy network so that they can evolve simultaneously. 

Training a deep neural network using reinforcement learning is time consuming because the only feedback for the training is delayed, sparse, and indirect in the form of rewards. Training a multi-agent reinforcement learning (MARL) model \cite{berner2019dota}\cite{vinyals2019grandmaster} is even more challenging due to the fact that agents’ decisions are not visible to one another. This lack of visibility reduces the predictability of the environment and makes it non-stationary. When a trainable message passing network is used to connect agents, things become even more complicated.  The additional trainable variables in the message network significantly increase the model’s complexity, prolong the training time, and escalate the chance of overfitting.

In this work, we accelerate the MARL by introducing another feedback channel that helps to learn a more efficient message passing network and a more effective representation of the environment. This consequently leads to better policy and faster convergence. In our belief-map assisted multi-agent system (BAMS)\footnote{The code is available at \href{https://github.com/qhuang18-97/BAMS.git}{Github}}, each agent is supplemented with a map decoder, which transforms its hidden state into a belief map, a neuro-symbolic representation of the agent’s knowledge of the global environment. This symbolic representation is simple, making it easy to obtain its corresponding ground-truth value. By comparing the belief map with the ground-truth map, the system receives an additional feedback that supervises the training process. During execution, the belief map provides a way to interpret the agent’s hidden state, which can further be used to explain the agent’s behavior.

To improve coordination among the agents and increase the efficiency of message retrieval, our message passing system incorporates gating and attention mechanisms. The attention model enables agents to differentiate important and irrelevant messages, while the gating removes the redundancy and saves communication power and bandwidth. 

We assessed the performance of the BAMS model using a multi-agent predator-prey game with and without obstacles. Centralized training and distributed execution are adopted in the experiments. The experimental results indicate that BAMS outperforms existing models, proving to be more fitting for large-scale environments with complex landscapes and providing more robust performance. 

\begin{figure*}[ht]
\begin{center}
\centerline{\includegraphics[width=\columnwidth/2*3]{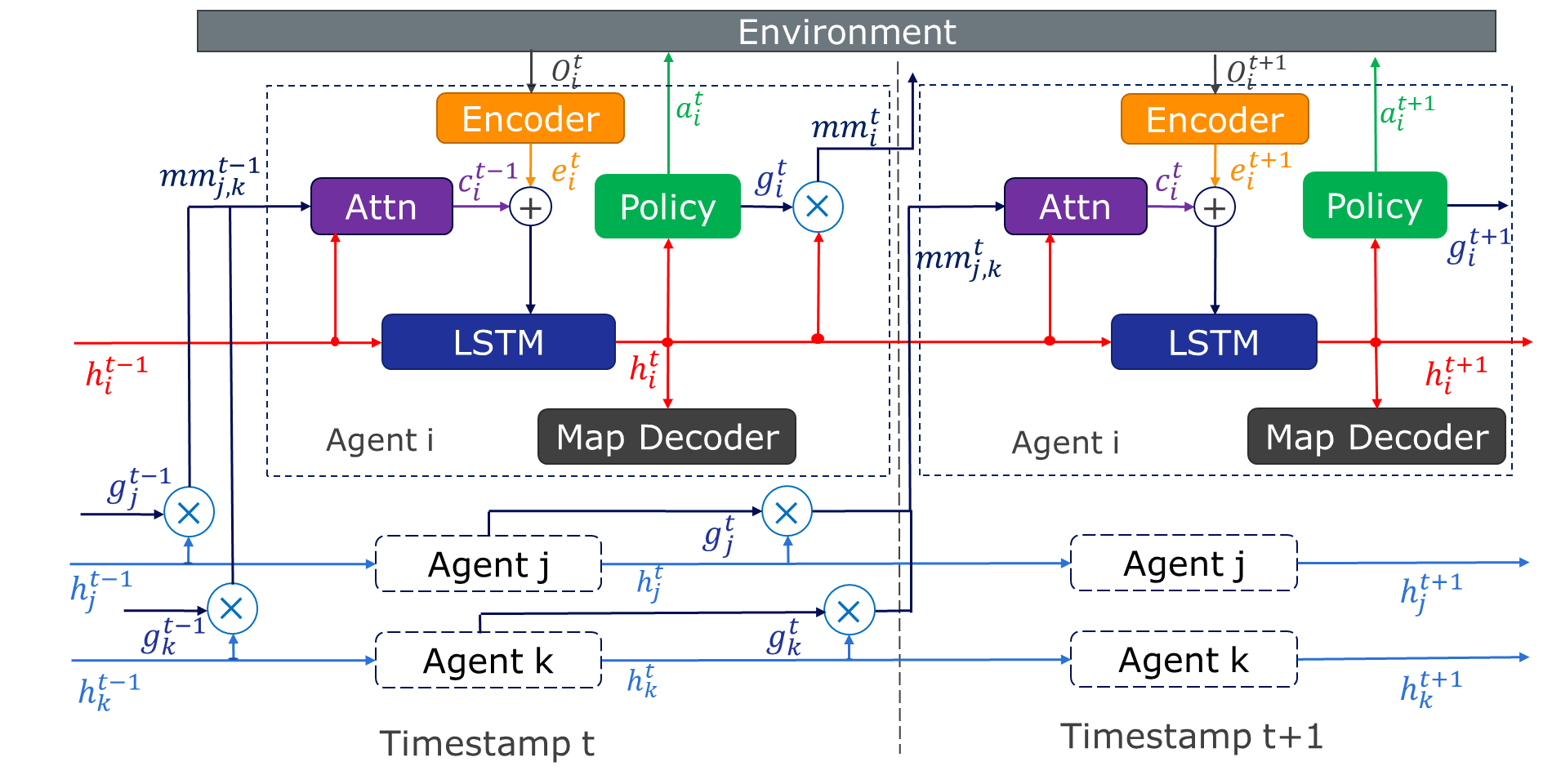}}
\caption{Architecture of BAMS model}
\label{figure1}
\end{center}
\vspace{-0.8cm}
\end{figure*}

The key contributions of this paper are summarized as follows:

\begin{itemize}
\item We proposed a belief-map assisted training mechanism that complements reinforcement learning with supervised information to accelerate training convergence. 
\item We proposed a belief-map decoder to reconstruct a neuro-symbolic map from the environment embedding to provide additional feedback during the training. The map transforms the hidden state of agents into a human-readable format, which significantly improves the interpretability of the agent’s decision-making process.
\item Agents trained using BAMS model communicate more effectively, catching the prey faster and being less susceptible to noises from redundant messages as the number of agents increases.
\item Simulation results show that agents with these enhancements can be trained effectively for operation in large and complex environments, reducing training time by an average of 66\% and improving overall performance by 34.62\%.
\end{itemize}

The rest of the paper is organized as follows. Section \ref{Works} introduces previous works related to communication in a multi-agent reinforcement learning system. Section \ref{Method} gives the details of our proposed method including the believe map decoder and attention model. The experimental results are given in Section \ref{Experiments} followed by the conclusions in Section \ref{Conclusion}.

\section{Motivations and Previous Work}
\label{Works}
We consider a fully cooperative multi-agent game as a \textit{decentralized partially observable Markov Decision Process(DEC-POMDP)} \cite{bernstein2002complexity}. 
DEC-POMDP is defined as a tuple $\left\langle \textit{N},\textit{S},\textit{P},\boldsymbol{\mathcal{R}},\boldsymbol{\mathcal{O}},\boldsymbol{\mathcal{A}},Z,\boldsymbol{\gamma}\right\rangle$, 
where $\textit{N}$ denotes the number of agents; 
$\textit{S}$ is a finite state space; 
$P(s'| s,a):S\times\boldsymbol{\mathcal{A}}\times S \to [0,1]$ stands for the state transition probabilities; 
$\boldsymbol{\mathcal{A}}=[\mathbf{A_1… A_N}]$ is a finite set of actions, where $\mathbf{A_i}$ represents the set of local actions $\mathbf{a_i}$ that agent $i$ can take; 
$\boldsymbol{\mathcal{O}}=[\mathbf{O_1}… \mathbf{O_N}]$ is a finite set of observations controlled by the observation function $Z:\textit{S}\times\boldsymbol{\mathcal{A}} \to \boldsymbol{\mathcal{O}}$; $\boldsymbol{\mathcal{R}}:\textit{S}\times\boldsymbol{\mathcal{A}} \to \mathbb{R}$ is the reward function; 
and $\gamma\in [0,1]$ is the discount factor.

According to the DEC-POMDP model, each agent \textit{i} takes an action $a_i$  based on its local observation $o_i$. 
When all agents applied their actions $[a_0,a_1,…,a_N]$ to the environment, 
the environment moves to a new state $s'$ and returns a joint reward $r$.
The MARL trains policies $\pi_i(a_i| o_i ): \boldsymbol{\mathcal{O}_i} \to \boldsymbol{\mathcal{A}_i}$,
$\forall i$, that maximizes the expected discounted reward $\mathbb{E}[\sum_{t=0}^{\infty}\gamma^t r^t]$, where $\gamma$ is the discount factor.

Sharing observation improves the performance and helps agents learn a better action policy. Efficient communication allows agents to obtain more information about the global environment and reduces the negative impact of partial observations. Previous research models a multi-agent communication system as a message passing graph neural network \cite{li2020graph}\cite{liu2020multi}, where each node in the graph represents an agent and each edge models a communication pathway equipped with message encoding and decoding. Different graph topologies have been studied \cite{sheng2022learning}, and recent works focus on improving the efficiency and reducing the cost of the communication using gated message passing \cite{mao2020learning}, attention \cite{geng2019learning}, schedule communication \cite{kim2019learning} and event/memory driven processing \cite{hu2021event}\cite{pesce2020improving} \cite{simoes2020multi}. 

The first study on learnable communication, known as RIAL and DIAL \cite{foerster2016learning}, developed a message passing network that generates message generation based on the agent’s local observation, action, and received messages. The message encoder is a multi-layer perceptron trained together with the policy network using reinforcement learning. CommNet \cite{sukhbaatar2016learning} includes a centralized communication channel into the network, which enhances  \cite{foerster2016learning} by maintaining a local hidden state in each agent using a recurrent neural network (RNN). The hidden state is determined by the sequence of local observations and received messages and is sent as the communication message to other agents. When multiple messages are received, the agent consolidates them using their average. 

Message gating \cite{jiang2018learning}\cite{singh2018learning} has been proposed as a binary action to dynamically block or unblock message transmission, thereby improving communication efficiency and conserving power and bandwidth. IC3Net \cite{singh2018learning}, an extension of CommNet \cite{sukhbaatar2016learning}, utilizes long short-term memory (LSTM) \cite{hochreiter1997long} to generate hidden states. Gated-ACML \cite{mao2020learning} performs message pruning before transmission. For both approaches, communication gating is optimized by the policy network using reinforcement learning.

Other studies \cite{jiang2018learning}\cite{liu2019multiagent}\cite{niu2021multi} have employed attention model to prioritize received messages so that agents can select useful features. ATOC \cite{jiang2018learning} applies attention to determine which agent to communicate with, and dynamically changes network structure accordingly by generated a directed graph. G2ANet \cite{liu2019multiagent} combines a hard attention and a soft attention as two stage attention model to process different incoming message from different agents. MAGIC \cite{niu2021multi} uses a multi-layout graph attention network among agents. However, it performs centralized communication and message processing. All messages are sent to a communication hub where they are consolidated using the attention model and then broadcasted to all agents.

TarMAC \cite{das2019tarmac} utilizes both gating and attention to enhance the communication efficiency. However, upon careful examination of its code, we found that an implementation error in the SoftMax function leads to unintended message leakage. If all agents gate their transmissions, the receiver may still receive this message. In other words, an agent $i$ can only gate its message to another agent $j$ if at least one other agent $k$,
$1 \leq\ k \leq\ \textit{N}$, 
$k \neq\ i$ or $j$,
decided to send its message to $j$ in the same cycle. As a result, agents must synchronize with each other regarding gating decisions during each cycle to determine whether to transmit messages. 

All the works mentioned above train the message passing network with the policy network using the game rewards as the feedback. This approach tends to have a slow convergence and the agents do not understand each other well. In this work, we proposed a belief-map assisted training method (BAMS) that significantly improves the training speed and quality for large and complex games. The agents trained using BAMS communicates more efficiently with fewer messages and better attentions. 

\section{Proposed Method}
\label{Method}

In this section, we present the structure and training of belief-map assisted multi-agent system (BAMS). Details of the BAMS are illustrated in Figure \ref{figure1}. For each agent $i$, the model comprises five major components: 

\begin{itemize}

\item Observation Encoder $E_i ()$: The observation encoder extracts key features from the agent’s local observation, which will later be combined with received messages and be used to update the hidden states.
\item Message Attention Module $A_i ()$: The attention module assigns weights to different messages to select relevant information. 
\item Hidden State Generator $lstm_i ()$: The hidden state generator is a Long Short-Term Memory (LSTM) that fuses the local observation and received messages into a feature vector $h_i$.
\item Policy Network $p_i ()$: The policy network is an actor-critic model that selects the best action for the local agent to maximize the overall system utility. In BAMS, the action consists of two parts, a discrete movement action $a_i$, which decides how agent moves to complete the game; and a binary communication action $g_i$, which decides whether the agent should broadcast its hidden state. The outgoing message $mm_i$ is the product of $g_i$ and $h_i$ as shown in Figure \ref{figure1}.
\item Map Decoder $D_i()$: The decoder reconstructs a neuro-symbolic belief map of the environment based on the hidden state of the local agent. The belief map represents agent’s knowledge of the global environment. It will be compared with the ground truth to provide additional feedback to assist the training.
 
\end{itemize}

\subsection{Hidden State Generation and Policy Network}

At each time step, every BAMS agent collects observations from its local sensor. The local observation for agent $i$ at time $t$ is denoted as $o_i^t$. Typically, the representation of $o_i^t$ is designed manually and tailored to the specific application. The agent then update its hidden state, which is maintained by an LSTM, using both local observations and the received messages as the following:
	\begin{equation} \label{eq:1}
	    h_i^{t+1},s_i^{t+1}=lstm_i (E_i (o_i^t ),c_i^t,h_i^t,s_i^t),
	\end{equation}
where $h_i^t$ and $s_i^t$ are hidden state and cell state at time $t$ of agent $i$, and $c_i^t$ is the aggregated feature extracted from the received messages using the attention model. $E_i (o_i^t )$ is the encoded observation.

Based on the hidden state, the agent chooses actions using a policy network $p_i ()$. The policy network follows the actor-critic model and comprises an actor network $\theta_i (h_i^t)$ and a critic network $V_i (h_i^t)$. The $\theta_i (h_i^t)$ is a one-layer fully connected network with an input of $h_i^t$. Its output has two components $a_i^t$ and $g_i^t$,
	\begin{equation}
	    a_i^t,g_i^t=\theta_i (h_i^t).
	\end{equation}
The vector $a_i^t$ represents the probabilities of the game actions available to the agent, i.e., the movement that the agent can make to complete the game. The variable $g_i^t$, which is either 1 or 0, represents the probability of the binary communication action, i.e., blocking or passing. At each step, the action was sampled according to the probability distribution.

\subsection{Message Passing Model}

Agents communicate their connected neighbors by sending messages. Following the approach used in TarMAC and IC3Net, we employ the hidden state as the communication message. The hidden state contains all the information that an agent requires to make local decisions. However, not all the information is useful to the agent’s neighbors. Furthermore, some of the information may overlap with previous messages from the same agent or messages sent by a nearby agent. To improve the efficiency of the communication network, the senders must reduce the number of redundant messages they send and the receivers must be able to extract useful information relevant to their own decision making. 


We implement the message gating at the sender side. The outgoing message $mm_j^t$ of agent $j$ is calculated as the product of $h_j^t$ and the binary gate action $g_j^t$. 
        \begin{equation}
            mm_j^t= h_j^t \times g_j^t.
        \end{equation}
After receiving messages $mm_j^t  (j\ne i)$ from neighbor $j$, agent $i$ aggregates the messages using an attention model, which is trained to maximize the reward from the game and minimize the loss of the belief-map construction. Considering the communication delay, agent $i$ uses gated message $mm_j^{t-1}$ send by agent $j$ in previous time step as the input of the key and value networks to generate $k_j^t$ and $v_j^t$ for time $t$. The query $q_i^t$ of the attention model is generated based on the agent’s local hidden state at current time step ($h_i^t$). 
\begin{equation}
    k_j^t=key(mm_j^{t-1} )	
\end{equation}
\begin{equation}
    v_j^t=value(mm_j^{t-1} ) 
    \end{equation}
    \begin{equation}
	q_i^t=query(h_i^t )  
 \end{equation}
 \begin{equation}
	\alpha_i^t=softmax\left[\frac{(q_i^{t^T} k_1^t)}{\sqrt{(d_k )}}…\frac{(q_i^{t^T} k_j^t)}{\sqrt{(d_k )}}…\frac{(q_i^{t^T} k_1^t)}{\sqrt{(d_k )}}\right]	
 \end{equation}
 \begin{equation}
	c_i^t=\sum\nolimits_{j=1}^N \alpha_i^t v_j^t 
\end{equation}
where $key()$, $value()$ and $query()$ are networks with one fully connected linear layer, $d_k$ is the dimensions of hidden state. $c_i^t$ is the aggregated feature vector that will be used to update the hidden state in Equation (\ref{eq:1}). 

\subsection{Map Decoder}

Instead of relying solely on the reward from the environment, additional channels of feedback information could be added to expedite the training process. In this work we assist the RL by using a decoded belief map. As the aggregation of past observations and incoming messages, an agent’s hidden state represents its knowledge of the environment. The more accurate this knowledge is, the better decision an agent can make. However, an agent’s hidden state is a feature vector that is not interpretable. The basic idea of BAMS is to decode the hidden state into a neuro-symbolic map that is human interpretable, allowing for the construction of the ground truth version of the map. By comparing the decoded map with ground truth map, we provide additional feedback to assist the training of the entire system.  

The map decoder $D_i(h_i^t)$ can be viewed as the inverse process of the observation encoder $E_i(o_i^t)$. The encoder $E_i(o_i^t)$ uses a Convolutional Neural Network (CNN) to extract the information. Therefore, we selected transposed CNN to decode the map. Both the observations and decoded maps are $m\times m$ gridded planes, where $m$ is the size of the environment. The status of each grid location is coded as a size $M$ vector, where $M$ represents the number of possible states of the grid. For example, in the predator-prey game, a grid can have 4 possible states that indicate whether it has been observed, is currently occupied by a predator, occupied by a prey, or occupied by an obstacle. These 4 states are not necessarily exclusive; hence each grid is encoded as a multi-hot vector with a size of $M$. Overall, both maps have dimension $M\times m\times m$. The observation map only contains information from the local agent, while the belief map should incorporate the information from all agents. 

\subsection{Loss Function}

In this work, we apply centralized training and distributed execution. All components in the BAMS are trained together.  

The training loss for each agent comprises two components, the $L_{map}$ and the $L_{RL}$, $Loss=\alpha L_{map} + \beta L_{RL}$, where $\alpha$ and $\beta$ are two hyper parameters. 
The map loss comes from comparing the decoded belief map $(b_i^t)$ and the ground truth map $\mathbb{G}_i^t$. Mean Squared Error (MSE) is used to calculate the loss, $L_{map}  = \sum_{t} MSE (\mathbb{G}_i^t  - b_i^t)$. During training, the central controller tracks the movement and status of all agents to generate ground truth map. The map loss is obtained in every time step $t$. Minimizing the map loss can help all agents converge to an effective communication protocol and efficient message processing. 

The RL loss is the error of the critic network,
\begin{equation}
    L_{RL}  = \sum\nolimits_{t} \lVert (r(h_i^{t},a_i^{t}) + \gamma \hat V(h_i^{t+1}) - \hat V(h_i^{t}) \rVert ^ 2
\end{equation}
where $r(h_i^{t},a_i^{t})$ is the reward of the entire system, and $\hat V()$ is the value estimation of the critic model. The actor network is updated using policy gradient:
\begin{equation}
    \nabla_{\theta} J(\theta)  = \sum\nolimits_{t} \nabla_{\theta} \log (p_{\theta}(a_i^t | h_i^t) [r(h_i^{t},a_i^{t}) + \gamma \hat V(h_i^{t+1}) - \hat V(h_i^{t})]
\end{equation}
where $p_{\theta}()$ is the prediction of the actor network. The BAMS model is updated using the average gradient of all agents.

\section{Experiments}
\label{Experiments}

For our experiments and evaluations, we utilized a classic grid-based predator-prey game \cite{liu2020multi}. It involves $N$ predators (agents) with limited vision ($v$) to explore an environment of size $m\times m$ to capture either a static prey or a moving prey. The value of $N$ ranges from 3 to 10, and $m$ ranges from 7 to 20, representing games with varying complexity. The environment is further divided into 2 categories, with obstacles and without obstacles.  

\subsection{Experiment Setting}

We trained our network using RMSprop \cite{ruder2016overview} with a learning rate of 0.001 and smoothing constant 0.97. The entropy regularization is used with coefficient 0.01. The hidden state size for LSTM is 64. For the attention model, the key ($k_j^t$) and query ($q_j^t$) have a dimension of 16 and the value ($v_j^t$) has a dimension of 64. 

The agents have limited observation capabilities. Specifically, each agent is only able to observe objects within a $3\times 3$ or $5\times 5$ area centered around itself. At each time step, an agent can choose from 5 possible actions: up, down, left, right, and stay. Additionally, all agents (predators) have a maximum step limitation, which varies according to the size of the environment. Prior to reaching the prey, an agent will receive a penalty $r_{searching}$ = -0.05 during each time step.  Once an agent reaches the prey, it will remain there and receives no further penalty. The game is considered as complete when all agents reach the prey within the maximum number of steps. The number of steps taken to complete the game serves as the performance metric.

We conducted a comparison of BAMS with 2 baseline models: TarMAC and IC3Net. I3CNet employs message gating while TarMAC employs both message gating and attention. To the best of our knowledge, these models have state-of-the-art performance while employing decentralized communication and decision-making. For TarMAC, regardless of the gating action, the message will be sent, together with the gating action. So we also implemented a variation of BAMS that removes the belief map decoder and conducts the training without the use of additional feedback. This reduced version of BAMS is referred to as BAMS-R. 


\begin{table*}[t]
\begin{center}
{\caption{Avg Steps \& Comm Rate for Simple Environments}\label{table1}}
\begin{tabular}{lcccccc}
\toprule
&\multicolumn{2}{c}{N=3, m=7, Max Steps = 20} &\multicolumn{2}{c}{N=5 m=12, Max Steps = 40} &\multicolumn{2}{c}{N=10, m=20, Max Steps = 80}    \\
& Avg steps & Comm rate & Avg steps & Comm rate & Avg steps & Comm rate \\ 
\midrule
Heuristic   & 14.56 & -    & 33.24 & -    & 68.90 & -         \\
IC3Net      & 12.48 & 0.60 & 32.90 & 0.39 & 73.82 & 0.60      \\
TarMAC      & 8.79  & 0.99 & 22.59 & 0.91 & 60.72 & 0.76      \\
BAMS-R      & 12.39 & 0.32 & 29.80 & 0.04 & 71.76 & 0.35      \\
BAMS        & \textbf{8.17}   & \textbf{21.64}  & \textbf{56.46}      \\ 
\bottomrule
\end{tabular}
\end{center}
\end{table*}

\begin{figure}[t]
\begin{center}
\centerline{\includegraphics[width=\columnwidth]{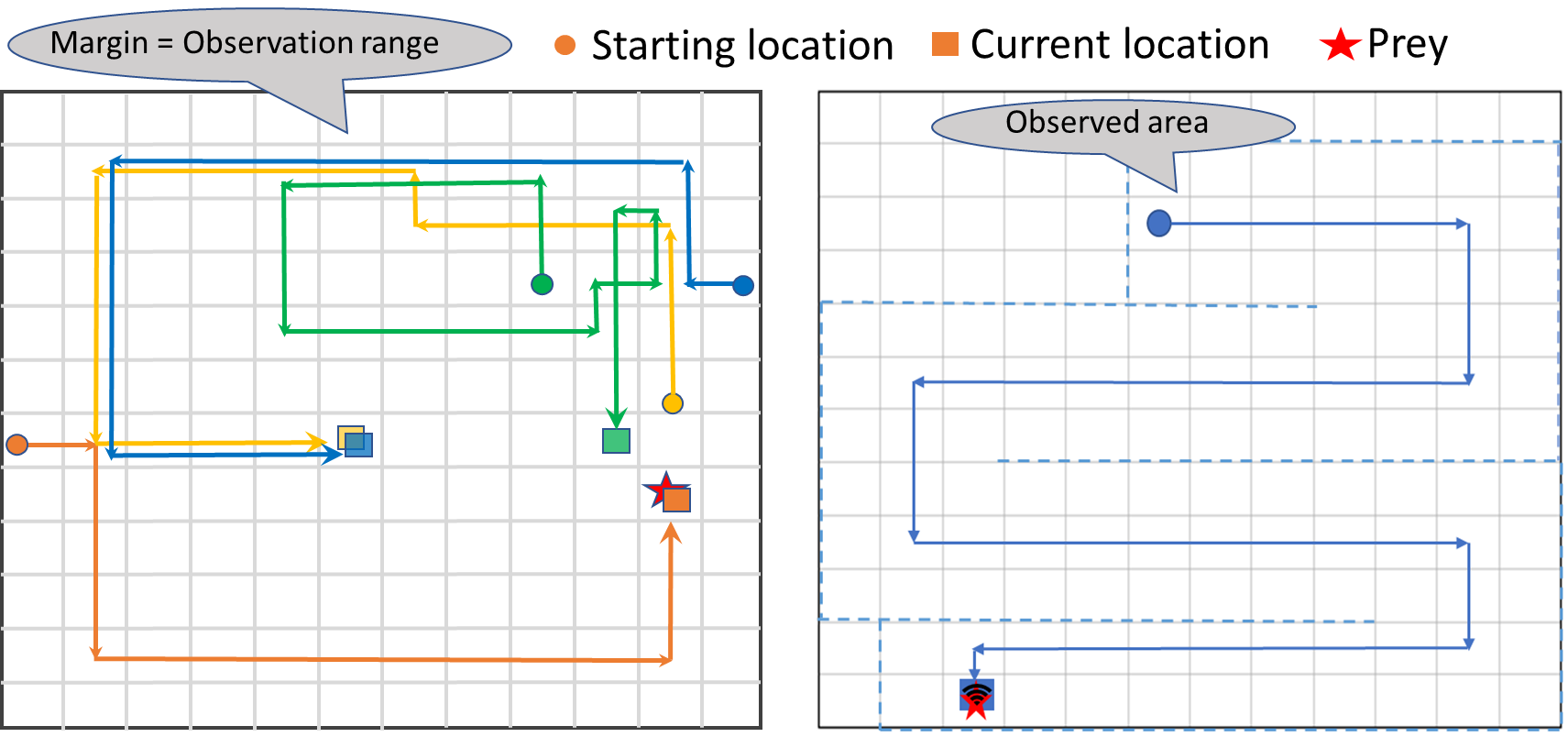}}
\caption{\textbf{Left} figure shows the example of 4 trajectories exhibiting keep the border within their observation range. \textbf{Right} figure shown the heuristic trajectory.}
\label{figure2}
\end{center}
\end{figure}

In addition to the aforementioned models, we implemented a heuristic rule-based algorithm. The algorithm directs the agents to explore the map from left to right, and top to bottom. after finishes exploring a row/column, an agent will move to the next row/column beyond its previous observation range. When it reaches the map’s edge, it will turn around and explore in the opposite direction. Once an agent has sighted the prey, it will transmit the prey’s location to all other agents, who will then take the shortest path to capture the prey. An example of the heuristic trajectory is shown as the right figure of Figure \ref{figure2}. 

\subsection{Experimental Results for Simple Environment}

The first experiment is carried out on simple environment without any obstacles. We discovered that agents developed significant levels of intelligence and mutual understanding, allowing them to complete the game with minimum communications. For example, all agents learned to explore the map by moving in a counterclockwise circle. Instead of exploring the entire map, agents circle a local region based on their initial position. Additionally, the agents tend to keep the border within their observation range while also staying as far from it as possible. These behaviors allow the agents to observe the maximum area while traveling the minimum distance. Figure \ref{figure2} presents an example of 4 trajectories exhibiting such behavior.

Table \ref{table1} compares the BAMS with four reference algorithms for games with different sizes when agents have $3\times 3$ vision. The column “comm rate” shows the average percentage of times an agent transmits its hidden state. The results indicate that BAMS takes fewer steps on average to complete the game than the other algorithms. Specifically, compared to IC3Net and the heuristic algorithm, BAMS completes the game with approximately 30\% fewer steps on average. Compared to TarMAC, BAMS completes the game with 6\% fewer steps. However, it should be noted that agents using TarMAC transmit their hidden state much more frequently. Moreover, as mentioned in Section \ref{Method}, agents in TarMAC must synchronize with each other about their gating decision at every time step, which incurs significant overhead. The comparison between BAMS and BAMS-R demonstrates that the use of belief-map assisted training leads to a 27\% reduction in the number of steps required to complete the game. As the map size increases, the communication rate of the BAMS agents reduces as the possibilities of encountering new events, such as observing another agent, the map edge, or the prey, decreases. In other words, the agents spend most of the time moving straight ahead. 

\begin{figure}[t]
\begin{center}
\centerline{\includegraphics[width=\columnwidth]{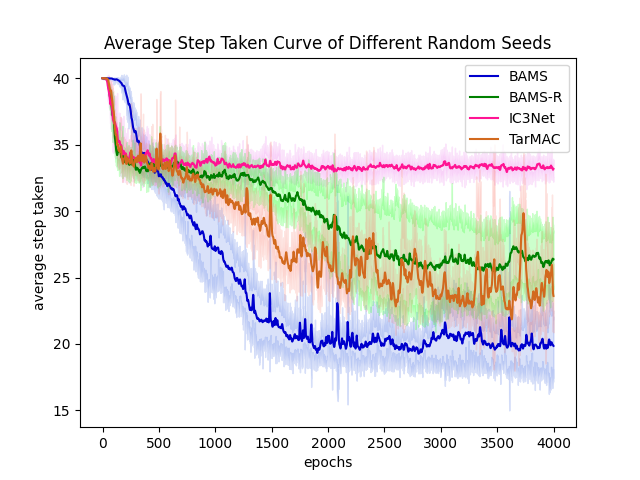}}
\caption{The convergence curve of four methods’ average step taken in random seeds under simple $12 \times 12$ environment.}
\label{figure3}
\end{center}
\end{figure}

Figure \ref{figure3} compares the convergence speed of BAMS with BAMS-R, Tarmac and IC3Net. The results indicate that, IC3Net has the fastest convergence due to its relatively simpler architecture that does not employ an attention mechanism in message processing. However, for the same reason, it also has the worst performance. On average, BAMS improves the convergence by 66\% compared BAMS-R. This improvement can be attributed to the additional feedback from the belief map, which provides a more consistent relationship among hidden state, action, and reward, resulting in faster learning with fewer iterations. Even TarMAC sends out messages every time step, our BAMS still beat the convergence of Tarmac. It should be noted that this feedback is only available during the training as no ground truth map is available during the execution. Nevertheless, the decoded belief map can provide a visualization of the agent’s hidden state and hence can be used to interpretate the agent’s decision-making process. 

\begin{figure*}[ht]
\begin{center}
\centerline{
\subfigure[Step 1 Ground Truth Map]{\includegraphics[width=\columnwidth*2/3]{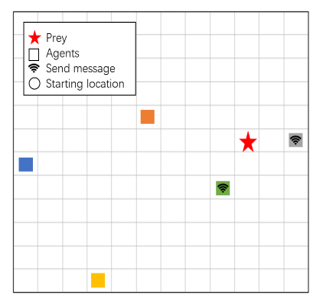}\label{figure4a}}\subfigure[Step 1 Decoded Map]{\includegraphics[width=\columnwidth]{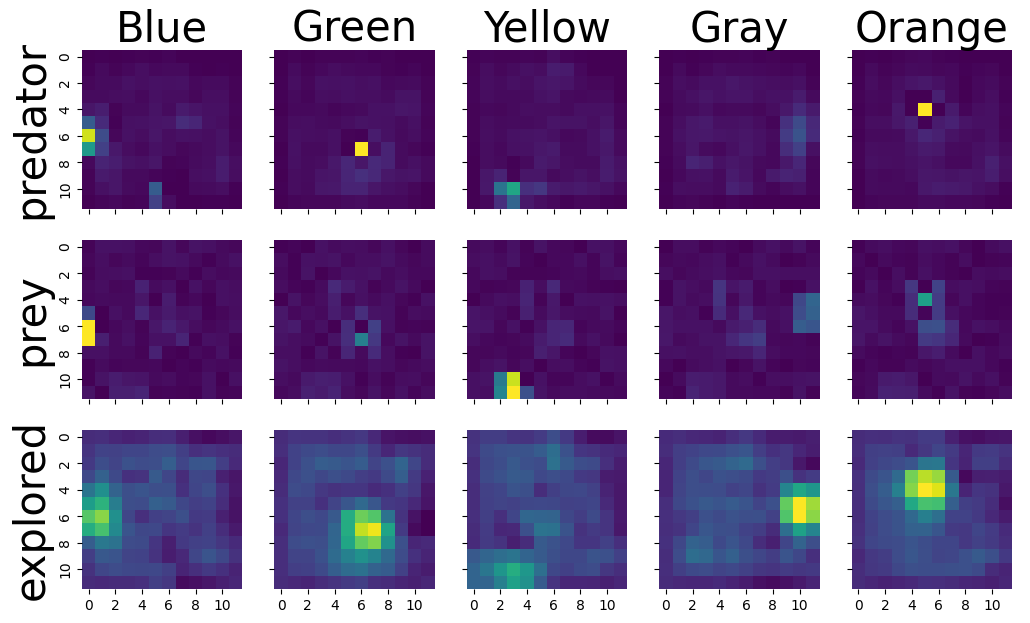}\label{figure4b}}
}
\centerline{
\subfigure[Step 3 Ground Truth Map]{\includegraphics[width=\columnwidth*2/3]{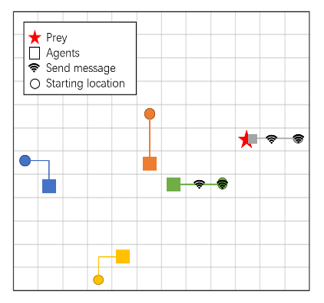}\label{figure4c}}\subfigure[Step 3 Decoded Map]{\includegraphics[width=\columnwidth]{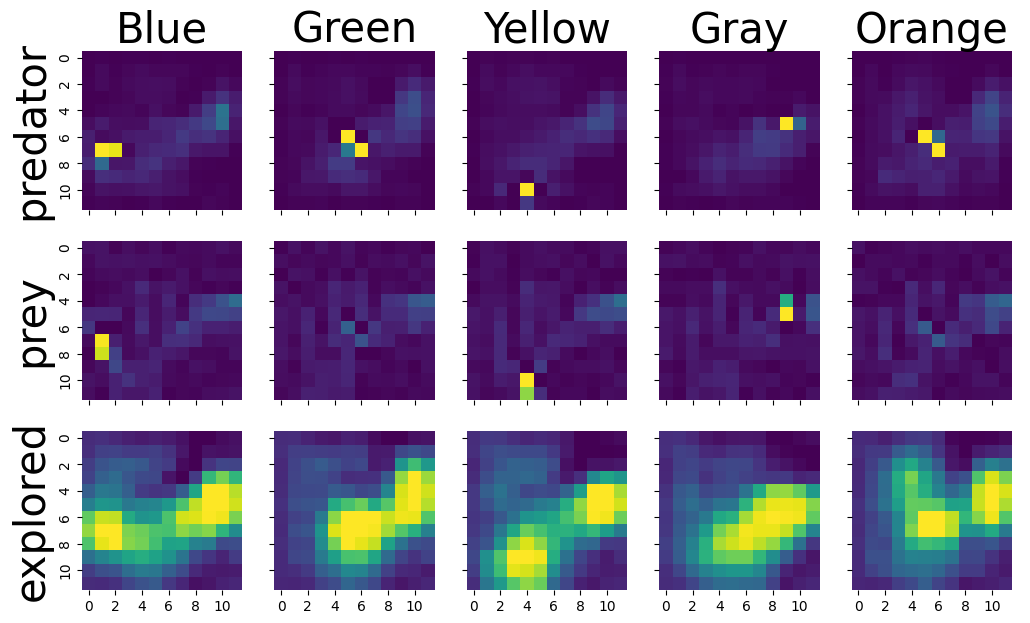}\label{figure4d}}
}
\raggedright
\caption{Visualization in Simple $12 \times 12$ environment of  Step 1 and Step 3. \textbf{Left} grid figures (a) (c) is ground truth map shows the trajectory of agents. Square represents agent and star represents prey. Circle represents the starting location of agent, and the Wi-Fi icon represents that agent sent out a message on that step. \textbf{Right} heatmap figures (b) (d) give the visualized belief map of agents. Brighter grids indicate higher possibility that the grids are taken by agents, prey, or explored.}
\label{figure4}
\end{center}
\end{figure*}

\begin{table*}
\begin{center}
{\caption{Scalability Analysis of Model with Varying Numbers of Agents}\label{table2}}
\resizebox{\columnwidth * 2}{!}{
\begin{tabular}{lcccccccccc}
\toprule
&\multicolumn{2}{c}{2} &\multicolumn{2}{c}{5} &\multicolumn{2}{c}{7} &\multicolumn{2}{c}{10} &\multicolumn{2}{c}{15}  \\
& Avg steps & Comm rate & Avg steps & Comm rate & Avg steps & Comm rate & Avg steps & Comm rate & Avg steps & Comm rate \\ 
\midrule
Heuristic   & 33.36 & -    & 27.40  & -    & 25.94 & -    & 20.47  & -    & \textbf{15.56}  & -    \\
IC3Net      & 29.34 & 0.43 & 32.90  & 0.39 & 33.13 & 0.42 & 34.91  & 0.42 & 35.74  & 0.39 \\
TarMAC      & 23.03 & 0.96 & 22.59 & 0.91  & 23.67 & 0.81 & 24.55 & 0.75 & 24.79 & 0.63     \\
BAMS-R      & 28.53 & 0.04 & 29.80  & 0.04 & 32.55 & 0.04 & 33.77  & 0.05 & 34.82  & 0.06 \\
BAMS        & \textbf{23.04} & 0.31 & \textbf{21.64} & 0.27 & \textbf{19.32} & 0.28 & \textbf{18.76} & 0.29 & 18.88 & 0.30\\ 
\bottomrule
\end{tabular}
}
\end{center}
\end{table*}

Figure \ref{figure4b} and Figure \ref{figure4d} depict an example of the decoded believe map for five agents at the beginning of the game and at time step 3 of the game, respectively. The gridded map shows the location of the agents and the prey, and the location where the agent sends out a message. The 3 channels of the decoded map indicate the belief of the agents’ location, the prey’s location and the explored area. At the beginning of the game, all agents only have access of their local information. Interestingly, we found that the agents learned to be optimistic, as each agent believed that the prey was located nearby. The Gray agent reached the prey in step 2 and both Gray and Green agents sent out messages in steps 1 and 2. Therefore, at step 3, all agents updated their belief map to reflect the messages they received. In their prey location map, the areas around location $(5, 9)$ is highlighted, which reflects the correct prey location that they learned from the Gray agent. In their explored area map, the right side and center area of the map are highlighted, indicating the area that has been explored by the Gray and Green agents. The Green and Orange agents observed each other in step 3, resulting in the highlighting of each other's location in their location map. Interestingly, even though the Blue and Yellow agents did not send out any messages, the other agents still slightly highlighted the left and bottom sides of their explored area maps, as if they anticipated someone exploring this area. This suggests a type of mutual understanding without direct communication. 

 To test the robustness of the policies, we train the BAMS, BAMS-R and IC3Net model in an environment with 5 agents and test them in different environments with agent numbers varying from 2 to 15. The results are reported in Table \ref{table2}, where we also listed the performance of the heuristic algorithm as a reference. As we expected, for BAMS, the average number of steps needed to complete the game reduces as the number of agents increases. However, for IC3Net and BAMS-R, the trend goes in the opposite direction. As the number of agents increases, due to the increased number of messages, the agents have difficulty extracting useful information, resulting in an increased number of steps to complete the game. This experiment demonstrates that BAMS helps to train an effective message passing framework, allowing agents to perform better in the game.

\subsection{Experimental Results for Complex Environment}

In the second experiment, we tested our approach under a complex environment with obstacles. Each grid in the environment is encoded as multi-hot vector of size 4, which represents whether the grid is occupied by a predator, a prey, or an obstacle, and whether it has been observed. We fixed the environment size to be $12\times 12$. In each randomly generated training environment, there are 20 randomly placed obstacles

\begin{figure}[ht]
\begin{center}
\centerline{\subfigure[Simple $20\times 20$]{\includegraphics[width=\columnwidth/2, height=3.5cm]{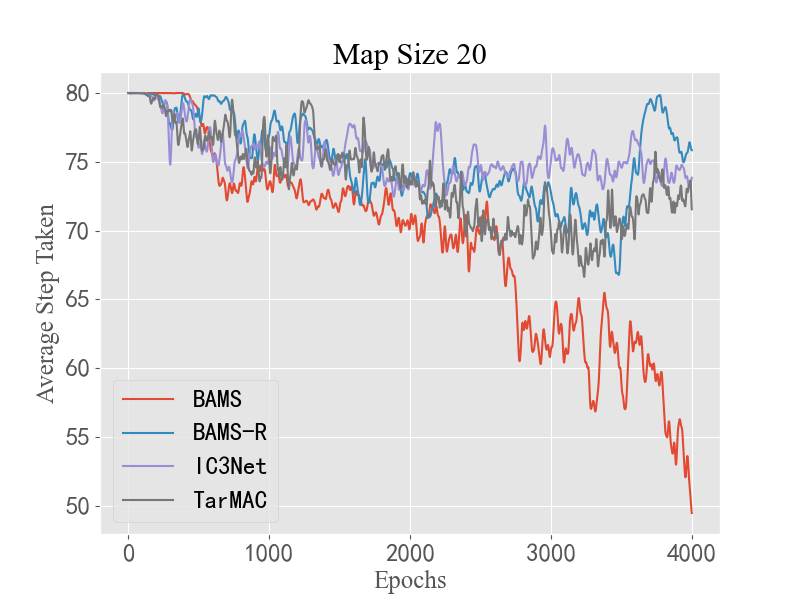}\label{figure5a}}\subfigure[Complex $12 \times 12$]{\includegraphics[width=\columnwidth/2, height=3.5cm]{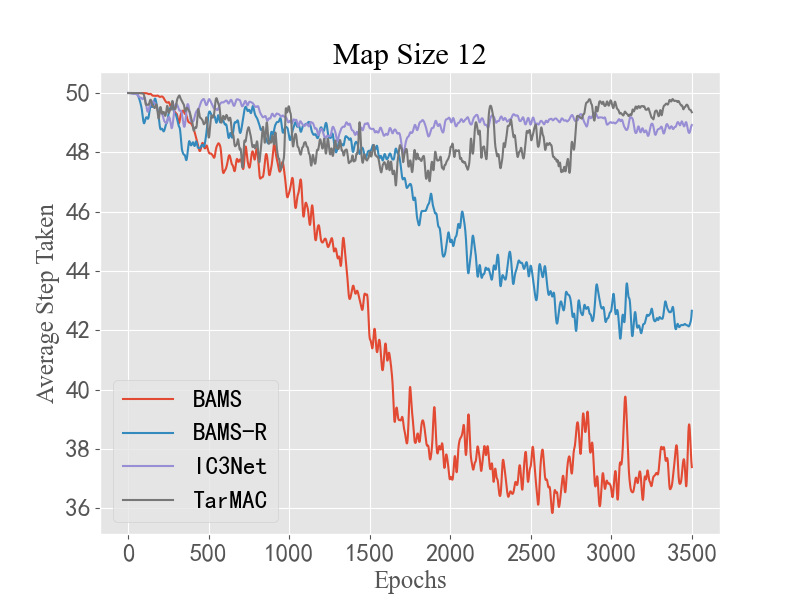}\label{figure5b}}}
\caption{Average Step Taken Comparison}
\label{figure5}
\end{center}
\end{figure}

Figure \ref{figure5b} compares the convergence speed of BAMS, Tarmac, IC3Net and BAMS-R. We can see that BAMS once again has the fastest convergence speed compared to the other models, completing the game with 3 fewer steps than the other two models in average. In comparison to Figure \ref{figure5a}, the performance of IC3Net, which does not employ attention to the received messages, deteriorates much faster than Tarmac, BAMS-R and BAMS. This means effective message passing network becomes increasingly important in a complex environment.

We also observed that as the environment becomes more complex, the performance of those models oscillates more significantly. This can be seen in Figure \ref{figure5a} and Figure \ref{figure5b} when the environment size is 20 or when obstacles are included. The reason for this is that in randomly generated large and complex environments, the difficulty level of the game can vary significantly. Factors such as the initial location of the agents and distribution of obstacles can affect the number of steps needed to complete the game. 

We further tested the model using testing environments with 10, 20 and 30 obstacles. We found that even though the network is trained with 20 obstacles, it was able to handle different environments. The performance of the three deep learning models in a complex environment is shown in Table \ref{table3}. In average BAMS reduces the number of steps by 23.6\% and 16.5\% compared to IC3Net and BAMS-R, respectively. 

\subsection{Experimental Results for Moving Prey}

We created a dynamic prey environment where the prey is able to move in order to evade capture by the agents. The prey has the same observation and action space as the agents.  When one or more agents are observed, the prey will move to the nearby grid that has the farthest Euclidean distance from the observed predators. In the case of a tie, the direction of movement is chosen randomly. The game ends when at least one agent successfully captures the prey. 

\begin{table}
\begin{center}
\caption{Avg Steps \& Comm Rate for Complex Environments
\label{table3}}
\resizebox{\linewidth}{!}{
\begin{tabular}{lcccccc}
\toprule
No. of obstacles&\multicolumn{2}{c}{10} &\multicolumn{2}{c}{20} &\multicolumn{2}{c}{30}    \\
& Avg steps & Comm rate & Avg steps & Comm rate & Avg steps & Comm rate \\ 
\midrule
IC3Net  & 45.39 & 0.53  & 48.56 & 0.54  & 49.37 & 0.57        \\
BAMS-R  & 39.43 & 0.062 & 44.78 & 0.073 & 46.92 & 0.076       \\
BAMS    & \textbf{31.80}  & 0.056 & \textbf{36.51} & 0.065 & \textbf{41.42} & 0.054       \\ 
\bottomrule
\end{tabular}
}
\end{center}
\vskip -0.1in
\end{table}

\begin{table}
\begin{center}
{\caption{Avg Steps \& Comm Rate for Moving Prey Environments}\label{table4}}
\resizebox{\linewidth}{!}{
\begin{tabular}{lcccc}
\toprule
Vision Size&\multicolumn{2}{c}{$3\times 3$} &\multicolumn{2}{c}{$5\times 5$}\\
& Avg steps & Comm rate & Avg steps & Comm rate\\ 
\midrule
BAMS & 32.47 & 0.94 & 35.64 & 0.96       \\
BAMS(pre-trained)     & 14.76  & 0.01 & 17.42 & 0.01       \\ 
\bottomrule
\end{tabular}
}
\end{center}
\end{table}

We further trained the BAMS model in a dynamic prey environment with a map size of 12 and observation range of $3\times 3$ and $5\times 5$. Two training strategies were tested. In the first approach, the BAMS model was trained from scratch in a moving prey environment. And in the $2^{nd}$ approach, we pre-trained the BAMS model in an environment with fixed prey, and then fine-tuned it in a moving prey environment. From Table \ref{table4} we can see that BAMS with pre-training outperforms the one without pre-training, reducing the average steps by more than 50\%. We also observed that BAMS without pre-train had a much higher communication rate. A possible explanation for the performance discrepancy is that the game with moving prey has two different goals: locating the prey and catching the prey. The former has a relatively stable environment, while the latter has a constantly changing environment as the prey is escaping. It is difficult to learn an effective communication strategy and an environment representation for both goals in one round. 

\section{Conclusion}
\label{Conclusion}

This paper proposes a novel training approach called belief map assisted training to improve the convergence and efficiency of multi-agent cooperative games with distributed decision-making. To overcome the issue of partial observation, attention-based inter-agent communication is adopted. The agents are trained to learn when to gate the message to save bandwidth and avoid interference with irrelevant information. We compared our approach with IC3Net and TarMAC in both simple and complex predator-prey environments. The experimental results show that our attention-based belief map can help the agents learn a better representation of the environment's hidden state and process messages effectively, leading to wiser decisions. Additionally, the belief map assisted training improves convergence speed and reduces the average number of steps needed to complete the game.

\clearpage

\ack This research is partially supported by the Air Force Office of Scientific Research (AFOSR), under contract FA9550-24-1-0078 and NSF under award CNS-2148253. 

The paper was received and approved for public release by AFRL on May $16^{th}$ 2023, case number AFRL-2023-2374. Any Opinions, findings, and conclusions or recommendations expressed in this material are those of the authors and do not necessarily reflect the views of AFRL or its contractors. 

\bibliography{ecai}
\end{document}